\documentclass[12pt]{article}

\usepackage{multicol}
\usepackage[pdftex]{graphicx}
\usepackage{booktabs}
\usepackage{amssymb,bm,mathrsfs,bbm,amscd}
\usepackage[tbtags]{amsmath}
\usepackage{lastpage}
\usepackage{subfigure,rotating,rotate}

\newcommand\pubnumber{SNSN-323-63\\UCHEP-11-01}
\newcommand\pubdate{\today}

\def\UC{Department of Physics, Mail Location \#0011, \\
University of Cincinnati\\
Cincinnati, Ohio, 45221-0011, USA}
\def\support{\footnote{Supported by the US National Science Foundation, under grant number PHY-0757876.
             \\ \quad  Representing the \babar collaboration.}}

\def\Title#1{\begin{center} {\Large #1 } \end{center}}
\def\Author#1{\begin{center}{ \sc #1} \end{center}}
\def\Address#1{\begin{center}{ \it #1} \end{center}}

\newcommand\pubblock{\rightline{\begin{tabular}{l} \pubnumber\\
         \pubdate  \end{tabular}}}
\newenvironment{Abstract}{\begin{quotation}  }{\end{quotation}}
\newenvironment{Presented}{\begin{quotation} \begin{center}
             PRESENTED AT\end{center}\bigskip
      \begin{center}\begin{large}}{\end{large}\end{center} \end{quotation}}

\input{babarsym.tex}

\newcommand{\sst}{\scriptscriptstyle}

\newcommand{\dzdzb}{\ensuremath{\Dz\Dzb}}
\newcommand{\beq}{\begin{equation}}
\newcommand{\eeq}{\end{equation}}
\newcommand{\bea}{\begin{eqnarray}}
\newcommand{\eea}{\end{eqnarray}}
\newcommand{\bean}{\begin{eqnarray*}}
\newcommand{\eean}{\end{eqnarray*}}
\newcommand{\CPV}{\ensuremath{C\!P\!V}}
\newcommand{\xd}{\ensuremath{x}}
\newcommand{\yd}{\ensuremath{y}}
\newcommand{\ycp}{\ensuremath{y_{C\!P}}}
\newcommand{\xp}{\ensuremath{x^{\prime}}}
\newcommand{\yp}{\ensuremath{y^{\prime}}}
\newcommand{\xpsq}{\ensuremath{x^{\prime 2}}}
\newcommand{\ypsq}{\ensuremath{y^{\prime 2}}}
\newcommand{\xpp}{\ensuremath{x^{\prime\prime}}}
\newcommand{\ypp}{\ensuremath{y^{\prime\prime}}}
\newcommand{\ifb}{\ensuremath{{\rm fb}^{-1}}}

\newcommand{\mco}{\multicolumn}
\newcommand{\kmpip}{\ensuremath{K^-\!\pi^+}}
\newcommand{\kppim}{\ensuremath{K^+\!\pi^-}}

\begin{document}
\begin{titlepage}
\pubblock

\vfill
\Title{CP Violation in Charm Mixing Results from Belle, BaBar and Tevatron}
\vfill
\Author{Brian Meadows\support}
\Address{\UC}
\vfill
\begin{Abstract}
Evidence from the \babar, Belle and CDF experiments for the phenomenon of \dzdzb\ oscillations is reviewed.  A summary is made of the current understanding of the parameters defining the mixing of mass eigenstates that give rise to the oscillations.  Results of searches for CP violation induced by mixing, are given, and estimates for the precision of measurements that can be expected in future experiments are made.
\end{Abstract}
\vfill
\begin{Presented}
CKM 2010, the 6th International Workshop on the CKM Unitarity Triangle, \\ University of Warwick, UK, 6-10 September 2010
\end{Presented}
\vfill
\end{titlepage}
\def\thefootnote{\fnsymbol{footnote}}
\setcounter{footnote}{0}
%


\section{Introduction}

\dzdzb\ oscillations arise from beating of two mass eigenstates, $D_1,~D_2$ that propagate differently in time, and are related to the instaneous flavour states by
\beq\label{eq:pq}
  \begin{array}{ll}
    \left.
    \begin{array}{rl}
     |D_1\rangle~=& p|\Dz\rangle\! +\! q|\Dzb\rangle \\
     |D_2\rangle~=& p|\Dz\rangle\! -\! q|\Dzb\rangle
    \end{array}
    \right\}
    ~~ |p|^2+|q|^2=1 ~~;~~ \arg\{q/p\}=\phi_{\sst M}.
  \end{array}
\eeq

Their amplitude and frequency depend upon the normalized differences
\bea\label{eq:xy}
  x = (m_1-m_2)/\Gamma &;&
  y = (\Gamma_1-\Gamma_2)/(2\Gamma) ~~;~
  \Gamma=(\Gamma_1+\Gamma_2)/2,
\eea
in the $D_1$ and $D_2$ masses and decay rates, $m_{1,2}$ and $\Gamma_{1,2}$, respectively.

The leading term in the standard model, SM, contributing to $\Delta C=2$ 
transitions ($C$ is the charm quantum number) is a box diagram with $W$ 
and $d$-type quarks.  This predicts very little mixing ($\xd<10^{-5}$)
\cite{Datta:1984jx},
because the SM was designed 
\cite{Glashow:1970gm} 
to produce large cancellation, $\sim (m_s^2-m_d^2)/m_c^2$, between $s$ and $d$ quarks, and also because the Cabibbo-Kobayashi-Moskawa (CKM) coupling $V_{ub}$ is small.  New physics (NP) could lead to larger values for $\xd$ or $\yd$
\cite{Petrov:2006nc,Petrov:2011un,Burdman:2003rs}, 
and could also produce \CP violation (\CPV) in mixing.  However, long range SM
effects from real, $\Delta C=1$ intermediate hadron states have also been suggested
\cite{Wolfenstein:1985ft}
to contribute significantly to both $\xd$ and $\yd$.  Theoretical uncertainty in ways to estimate these sums exists
\cite{Donoghue:1985hh},  
and values for $\xd$ and $\yd$ over a wide range ($10^{-7}$-$10^{-2}$) have been quoted
\cite{Petrov:2006nc,Falk:2004wg}.  
The present experimental results fit well into the higher range, but
the possibility that NP may be involved cannot be ruled out.  
Theoretical consensus is that NP is neither required nor ruled out by 
present measurements.  It is mostly agreed that observable \CPV\ at present experimental sensitivities would signify unambiguous evidence for NP were it observed
\cite{Petrov:2011un}.


\subsection{\boldmath{\Dz} decays}
\label{sec:d0decays}

Measurements of \dzdzb\ mixing have been based on observed time-dependences for decays of \Dz\ to final states $f$ accessible to either \Dz\ or \Dzb\ (\eg\ \Kp\!\!\Km, \Kp\!\pim, \Kp\!\pim\!\piz, \etc
\footnote{Unless explicitly stated otherwise, charge conjugate
states are implied.}).
In such decays, mixing and direct decay interfere.
%
%
In the abscence of \CPV, the number $N(t)$ of \Dz\ remaining at time $t$ is given, to second order in $x$ and $y$ by
\begin{align}
\label{eq:nonexponential}
     N(t) = N(0) e^{-\Gamma t} \times
            \left[1+|\lambda_f|(y\cos\theta_f-x\sin\theta_f)(\Gamma t)+
            \frac{x^2+y^2}{4}|\lambda_f|^{2}(\Gamma t)^2\right].
\end{align}
%
The first term in square parentheses corresponds to direct decay $\Dz\to f$ (amplitude ${\cal A}_f$).  The term quadratic in $t$ represents mixing ($\Dz\to\Dzb$) followed by decay $\Dzb\to f$ (amplitude $\bar{\cal A}_f$).  The middle term, linear in $t$, is due to the interference between these processes.  The parameter $\lambda_f$ is defined as
\begin{align}
  \lambda_f         &=
  \left(q\bar{\cal A}_f\right)/\left(p{\cal A}_f\right) ~~;~~
  \arg\{\lambda_f\}  =
  \theta_f   = \phi_{\sst M}+\phi_f+\delta_f.
\end{align}
These amplitudes $\bar{\cal A}_f$ and ${\cal A}_f$ have relative weak (strong) phases $\phi_f$ ($\delta_f$).

If \CP\ is conserved in mixing, $p=q=1/\sqrt{2}$ and $\phi_{\sst M}=0$.
If it is conserved in decay, then $\phi_f=0$.  In the special case that $f$ is a \CP-eigenstate, then $\phi_f=\delta_f=0$ and $\lambda_f=\pm e^{i\phi_{\sst M}}$.  In most measurements so far, \CPV\ is ignored and then $\theta_f=\delta_f$.

The strong phase difference $\delta_f$ is, in general, unknown and means that, for many analyses, only the rotated parameters
\beq
  \label{eq:xpyp}
     \xp  =  x\cos\delta_f + y\sin\delta_f ~~;~~
     \yp  =  y\cos\delta_f - x\sin\delta_f 
\eeq
can be determined.


\subsection{\label{sec:flavor}Flavor tagging and decay lengths}

Identifying the flavor (\Dz\ or \Dzb) of a neutral $D$ at birth is sometimes 
required.  When this is so, each $\Dz$ candidate is required to come from a 
$D^{*+}\to\Dz\pi_s^+$ decay, where the sign of the low momentum pion,  $\pi_s$, tags
the $D$ flavor.  $D^{*}$'s are identified by the required peak in the 
distribution of $\Delta M$, the difference between the invariant mass $M$ 
of the \Dz\ daughters and that of the $\Dz\pi_s^+$ system.  This peak also
serves to improve on signal selection and background rejection.

Decay length distributions for \Dz's from $B$ decays differ from those 
produced directly.  In the Belle and \babar\ analyses, \Dz's from $B$ are 
removed by selecting center of mass momentum of the \Dz\ above the kinematic
limit ($\sim$2.5~\gevc).  In the CDF data, the two sets of decays are 
distinguished by the distributions of their impact parameters.


\section{\label{sec:tauratio}\boldmath
Evidence for mixing in decays to \CP\ eigenstates}

The Belle collaboration first reported evidence for mixing with $3.2\sigma$ 
significance in decays of \Dz\ to \CP\ eigenstates $h^-h^+$ ($h=\pi$ or $K$)
\cite{Staric:2007dt}.
For such decays, if \CP\ is conserved,  mean decay times computed from 
Eq.~(\ref{eq:nonexponential}) $\tau$ are related to decays to non-\CP\ states
such as $\Dz\to\Km\pip$
\footnote{Here, as in all cases unless stated otherwise, the first uncertainty is statistical and the second is systematic.}
by
\beq
  \label{eq:ycp}
  \ycp \approx y = \frac{\tau(\Dz\to\kmpip)}{\tau(\Dz\to h^+\!h^-)}-1
\eeq
Using separate, flavor-tagged samples of \Dz\ and \Dzb, Belle also measured the \CP\ asymmetry:
\beq
  \label{eq:atau}
  A_{\tau} = \frac{\tau(\Dz\to h^+\!h^-)-\tau(\Dzb\to h^+\!h^-)}
                  {\tau(\Dz\to h^+\!h^-)+\tau(\Dzb\to h^+\!h^-)}
           = 0.010\pm 0.300\pm 0.150\%,
\eeq
consistent with zero.

This evidence was confirmed by the \babar\ collaboration in two independent 
measurements using flavor-tagged \Dz\ decays to $\Kp\!\!\Km$ and $\pip\!\pim$
\cite{Aubert:2007en}
and to a large, disjoint, untagged $\Kp\!\!\Km$ sample
\cite{Aubert:2009ck}
with a combined significance of $4.1\sigma$.  The \CP\ asymmetry $A_{\tau}$ 
for the \babar\ tagged decays was $0.260\pm 0.360\pm 0.080$\%, also consistent with zero.

These results, with other less significant ones have been combined
\cite{HFAG:2010qj}
to give a mean value $\ycp=(1.107\pm 0.217)\times 10^{-2}$, $5.0\sigma$ from the no mixing value zero.


\section{\label{sec:ws}Evidence for mixing in ``wrong sign" decays}

``Wrong sign" (WS) decays $\Dz\to\Kp\pim(\piz)$ are sensitive to mixing.  Here, direct decay is doubly Cabibbo suppressed (DCS) and the three terms in Eq.~(\ref{eq:nonexponential}) are all comparable.  For Cabibbo-favoured (CF) ``right-sign" (RS) decays $\Dz\to\Km\pip(\piz)$, however, the first term dominates and decays have an exponential distribution.  Experimentally, decay time distributions for both rare WS and copious RS events are fit simultaneously to determine the time resolution parameters (same for both) and mixing parameters (only observable in the WS sample).  For $\Dz\to\Kp\pim$ decays, $\delta_f$ is unknown and only rotated parameters \xpsq\ and \yp\ can be measured.

The \babar\ collaboration was first to report evidence for \Dz\ oscillations
\cite{Aubert:2007wf}
in a large sample of WS decays $\Dz\to\kppim$.  In a challenging and careful analysis of the decay time distribution in which the time resolution was only slightly less than the oscillation period, they reported a $3.9\sigma$ deviation.  In an earlier analysis of this channel, Belle
\cite{Zhang:2006dp}
had seen only a $2.0\sigma$ effect
\footnote{This analysis was even more sensitive than that of \babar, but the central values for \xpsq\ and \yp\ were closer to zero.}.
Subsequently, the evidence for mixing was confirmed by CDF
\cite{Aaltonen:2007uc}
with a $3.8\sigma$ significance.  Results from these three experiments are summarized in Table~\ref{tab:kpimix}.
\begin{table}[htb]
\begin{center}
\caption{Rotated mixing parameters $(\xpsq, \yp)$ in units of $10^{-3}$
         from fits to WS $\Dz\to\Kp\pim$ decays described in the text.
         The significance for mixing is obtained from 2-dimensional
         likelihood contours between the highly correlated \xpsq\ and \yp\
         values, that are computed to include systematic effects.  The                  significance is taken from the contour on which the no mixing
         point ($\xpsq=\yp=0$) lies.  The table includes \CP\ asymmetries
         for the DCS/CF ratio $R_{\sst D}=|\lambda|^{-2}$ for \Dz\ and
         $\bar R_{\sst D}=|\bar\lambda|^{-2}$ for \Dzb, and mixing rates
         $R_{\sst M}=\xpsq+\ypsq$ for \Dz\ and $\bar R_{\sst M}=
         \bar\xpsq+\bar\ypsq$ for \Dzb.}
\label{tab:kpimix}
\begin{tabular}{lccc}
\toprule
   Parameter & \babar & CDF & Belle \\
\toprule
  \xpsq\                   &
   $-0.22\pm 0.37$         &
   $-0.12\pm 0.35$         &
   $ 0.18^{+0.21}_{-0.23}$ \\
  \yp\                     &
   $ 9.7\pm 5.4$           &
   $ 8.5\pm 7.6$           &
   $ 0.6^{+4.0}_{-3.9}$    \\
  Mixing significance      &
   $ 3.9\sigma$            &
   $ 3.8\sigma$            &
   $ 2.0\sigma$            \\
   $a_{\sst D}=(R_{\sst D}-\bar R_{\sst D})/(R_{\sst D}+\bar R_{\sst D})$
                           &
   $ -21\pm 54$            &
                           &
   $  23\pm 47$            \\
   $a_{\sst M}= (R_{\sst M}-\bar R_{\sst M})/(R_{\sst M}+\bar R_{\sst M})$
                           &
   $  $                    &
                           &
   $ 670\pm 1200$          \\
\bottomrule
\end{tabular}
\end{center}
\end{table}
%


\subsection{\label{sec:k2pi}\boldmath Mixing in $\Dz\to\kppim\!\piz$ decays}

The \babar\ collaboration
\cite{Aubert:2008zh}
also reported evidence for mixing from a time-dependent Dalitz plot (DP) analysis of WS decays to the three-body system having an additional $\piz$ meson.  This system is similar to the two-body one, except that the final state $f$ is now a point in the $\kppim\!\piz$ phase space, specified by its DP coordinates $s_0$ and $s_+$, the squares of invariant masses of $\kppim$ and $\Kp\!\piz$ systems, respectively.

The expected time-dependence is also given by Eq.~(\ref{eq:nonexponential}), with the parameter $\lambda_f$ and its phase at each $f$ given by the ratio of decay amplitudes $\bar{\cal A}(s_0,s_+)$ for CF decay ($\Dzb\to\kppim\!\piz$) and ${\cal A}(s_0,s_+)$ for direct DCS decay ($\Dz\to\kppim\!\piz$) at that point.

In the \babar\ analysis, approximately 660K RS ($\kmpip\!\piz$) and 3K WS
($\kppim\!\piz$) flavor-tagged decays are  extracted from a 384~\ifb\ sample of \epem\ interactions.  Parameters for separate models $\bar{\cal A}_f$ and ${\cal A}_f$ for CF and DCS amplitudes, respectively, in each case a linear combination of Breit-Wigner (BW) amplitudes describing their different $K\pi$ and $\pi\pi$ resonance structures, are determined from simultaneous fits to the DP's for these samples
\footnote{Since the RS sample is dominated by CF decays, a time-integrated fit is made to determine $\bar{\cal A}_f$.  A time-dependent fit to the WS sample determines parameters for ${\cal A}_f$ and \xpp\ and \ypp\ values.}.
Assuming \CP\ is conserved, a model for the time-dependent variations in $\delta_f$ from point to point in the DP are thus obtained, but an unknown overall phase $\delta_{K\pi\pi}$ allows only rotated coordinates \xpp\ and \ypp, similar to \xp\ and \yp\ in Eq.~(\ref{eq:xpyp}), to be measured.

Values obtained for $\xpp$ and $\ypp$ are given in Table~\ref{tab:xppypp}.  A confidence level test, similar to that used in the WS $\kppim$ case indicates evidence for mixing at the $3.1\sigma$ level.  The major systematic uncertainties are associated with the assumptions in the BW model and in the description of the large background under the WS events.  Estimates of these are included in the computation of the confidence level.
\begin{table}[htb]
\begin{center}
\caption{Rotated mixing parameters \xpp\ and
            \ypp\ from fits to \babar\ data described in
            the text.  The first error is statistical and the
            second that attributed to systematic effects.}
\label{tab:xppypp}
\begin{tabular}{lcc}
   Sample & \xpp~(\%) & \ypp~(\%) \\
\hline
  \Dz\ and \Dzb\ &
   $ 2.61^{+0.57}_{-0.68}\pm 0.39$      &
   $-0.06^{+0.55}_{-0.64}\pm 0.34$      \\
  \Dz\ only      &
   $ 2.53^{+0.54}_{-0.63}\pm 0.39$      &
   $-0.05^{+0.63}_{-0.67}\pm 0.50$      \\
  \Dzb\ only     &
   $ 3.55^{+0.73}_{-0.83}\pm 0.65$      &
   $-0.54^{+0.40}_{-1.16}\pm 0.41$      \\
\hline
\end{tabular}
\end{center}
\end{table}

The fit procedure is repeated separately for $\Dz$ and $\Dzb$ samples to obtain values for $\xpp$ and $\ypp$ also listed in Table~\ref{tab:xppypp}, indicating no evidence for \CPV.


\section{\boldmath Decays $\Dz\to\KS\pi\pim$ and $\Dz\to\KS\Kp\Km$}

Decays to self-conjugate systems (sum of \CP-even and \CP-odd states) allow direct measurement of \xd\ and \yd\ since the strong phase difference between \Dz\ and \Dzb\ decays is zero.  Using these channels, it is also possible to obtain values for the \CPV\ parameters $|q/p|$ and $\phi_{\sst M}$.

This was first exploited for the $\Dz\to\KS\pip\pim$ mode by CLEO
\cite{Asner:2005sz}
using a 9~\invfb\ data sample and obtaining only an upper limit on mixing.  The Belle collaboration repeated the analysis
\cite{Abe:2007rd}
with a 540~\ifb\ sample to obtain central values for \xd, \yd, $|q/p|$ and $\arg\{q/p\}$.  More recently, \babar\ has used both $\Dz\to\KS\pi\pim$ and $\Dz\to\KS\Kp\Km$ modes for a 486.5~\ifb\ sample to obtain the most precise results.  Parameters from all three experiments are summarized in Table~\ref{tab:kshh}.
\begin{table}[htb]
\begin{center}
\caption{Mixing parameters from fits to $\Dz\to\KS h^+h^-$ ($h=\pi$ or $K$) decays.
         The first uncertainty is statistical, the second is from systematic
         effects.  A third uncertainty comes from ambiguities in the choice of model
         used to describe the decay amplitudes.}
\label{tab:kshh}
\begin{tabular}{|l|r@{=}lr@{=}l|}
\hline
\\[-6pt]
    \bf Experiment
     & \mco{4}{c}{\bf Mixing Parameters} \\[0pt]
\hline
     CLEO 2.5 (9 \ifb)
     & $\xd$&$(~1.9_{-3.3}^{+3.2}\pm 0.4\pm 0.4)$\% 
     & $\yd$&$(-1.4\pm 2.4\pm 0.8\pm 0.4)$\% 
\\[1pt]
     \hline
 
     BELLE (540 \ifb)
     & $\xd$&$(0.81\pm 0.30_{-0.07~-0.16}^{+0.10~+0.09})$\% 
     & $\yd$&$(0.37\pm 0.25_{-0.13~-0.08}^{+0.07~+0.07})$\% 
\\[1pt]
     (Allowing~ \CPV)
     &  $|q/p|$&$0.86_{-0.29~-0.03}^{+0.30~+0.06}\pm 0.08$ 
     &  $\phi_{\sst M}$&$(-14_{-18~-3~-4}^{+16~+5~+2})^{\circ}$ 
\\[1pt]
     \hline
     \babar (486.5 \ifb)
     & $\xd$&$(0.16\pm 0.23\pm 0.12\pm 0.08)$\% 
     & $\yd$&$(0.57\pm 0.20\pm 0.13\pm 0.07)$\% 
\\[1pt]
     \qquad(\Dz\ only)
     & $\xd^+$&$(0.00\pm 0.33)$\%
     & $\yd^+$&$(0.55\pm 0.27)$\%  
\\
     \qquad(\Dzb\ only)
     & $\xd^-$&$(0.33\pm 0.33)$\%
     & $\yd^-$&$(0.59\pm 0.28)$\% 
\\[1pt]
     \hline
  \end{tabular}
\end{center}
\end{table}

These results agree well and, since they do not depend upon any unknown strong phases, significantly affect the averages.  Central values for \xd\ and \yd, however, are such that the significance for mixing is small ($2.2\sigma$ for Belle and $1.9\sigma$ for \babar).

The decay amplitude models describing the DP distributions differ between the three $\KS\pip\pim$ analyses.  We note from Table~\ref{tab:kshh} that uncertainties from the decay amplitude model observed in \babar\ and Belle lead to irreducible uncertainties of order $10^{-3}$ in both \xd\ and \yd.  These, in turn, limit \CPV\ measurements to $\sim$25\% in $|q/p|-1$.


\section{\label{sec:future}Future Outlook}

The task of combining some 30 ``mixing observables", some of which are presented above, into values for the important physics parameters underlying them has been undertaken by the heavy flavor averaging group (HFAG
\cite{HFAG:2010qj}).
These are summarized in Table~\ref{tab:hfagmix}.  $\chi^2$ contours for this fit indicate that these central values are far (at least $10\sigma$) from the no mixing point $(\xd=\yd=0)$ and are within $1\sigma$ from the no \CPV\ point $(|q/p|=1, \phi_{\sst M}=0)$.
\begin{table}[htb]
\begin{center}
\caption{HFAG summary of mixing parameters from fits to 30 observables.}
\label{tab:hfagmix}
\begin{tabular}{|lc|lc|}
\toprule
   \xd\                           & $(6.3^{+1.9}_{-2.0})\times 10^{-3}$ &
   \yd\                           & $(7.5\pm 1.2)\times 10^{-3}$
  \\
  $|q/p|$                         & $0.91^{+0.18}_{-0.16}$   &
  $\phi_{\sst M}^{\circ}$         & $-10.2^{+9.4}_{-8.9}$
  \\
  $R_{\sst D}$                    & $(3.309\pm 0.081)\times 10^{-3}$ &
  $A_{\sst D}$                    & $(-19.2\pm 24)\times 10^{-3}$
  \\
  $\delta_{\sst K\pi}^{\circ}$    & $22.0^{+9.8}_{-11.2}$ &
  $\delta_{\sst K\pi\pi}^{\circ}$ & $19.3^{+21.8}_{-22.9}$
  \\
\bottomrule
\end{tabular}
\end{center}
\end{table}

Upgrades in both $B$ factories are planned that will increase event yields by a factor $\sim 100$.  New results from \dzdzb\ pair production at charm threshold from BES III are also anticipated to provide improved measurements of strong phase differences $\delta_f$ for $\Kp\pim$, $\Kp\pim\piz$ and also for individual points in various Dalitz plots that could be used for reducing uncertainties in decay amplitude models.  LHCb is running now, too, and will surely add to our knowledge and precision of mixing parameters very soon.

As we look ahead, we can consider how we might search for \CPV, a possible indicator for NP.  One way is to use the self-conjugate channels $\Dz\to\KS h^+h^-$ to directly measure $|q/p|$ and $\phi_M$.  From Table~\ref{tab:kshh} it is seen that these are limited by decay amplitude model uncertainty to $\sim 8$\% and $3^{\circ}$, respectively.  Another way that has been considered in detail by the \superb\ collaboration
\cite{O'Leary:2010af}
is to measure \CP\ asymmetries in the parameters \xd, \yd, \xp, \yp, \xpp\ or \ypp\ that each provide measurement of $|q/p|^2-1$.  To reach a precision in $|q/p|$ of 1\% requires precision in \xd\ or \yd\ of $\sim 10^{-4}$ that can be acheived only if decay model uncertainties can be reduced by a factor 10, as illustrated in Fig.~\ref{fig:bmfig}.  Asymmetries in these parameters in different channels at the $5\%$ level are also possible, and can be used to test for possible presence of direct \CPV.

\begin{figure}[htb]
\begin{center}
\begin{tabular} {ccc}
\resizebox{16cm}{!}{
  \includegraphics[height=0.15\textwidth]{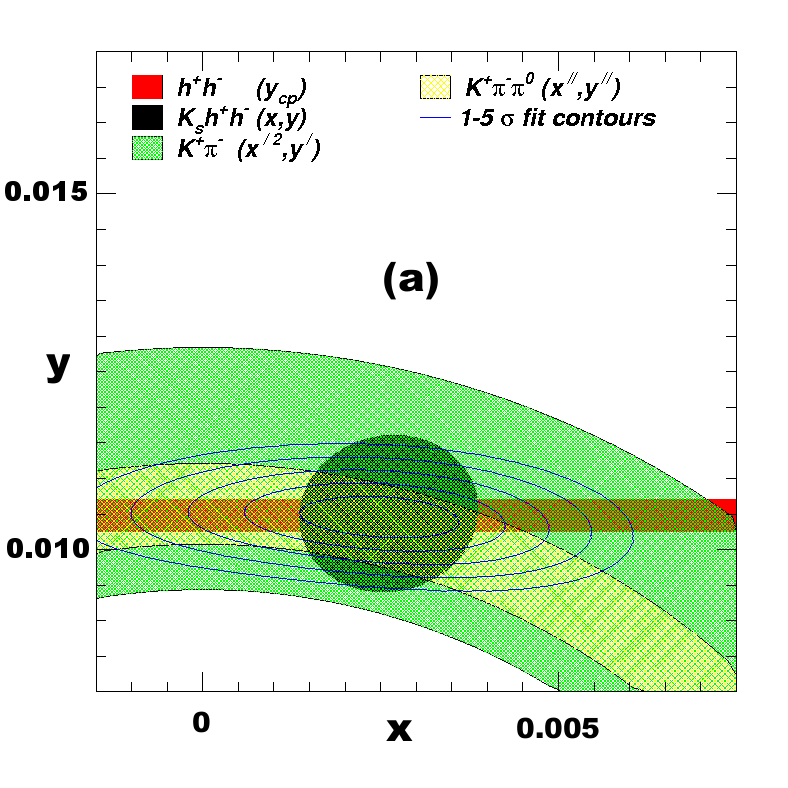}
  \includegraphics[height=0.15\textwidth]{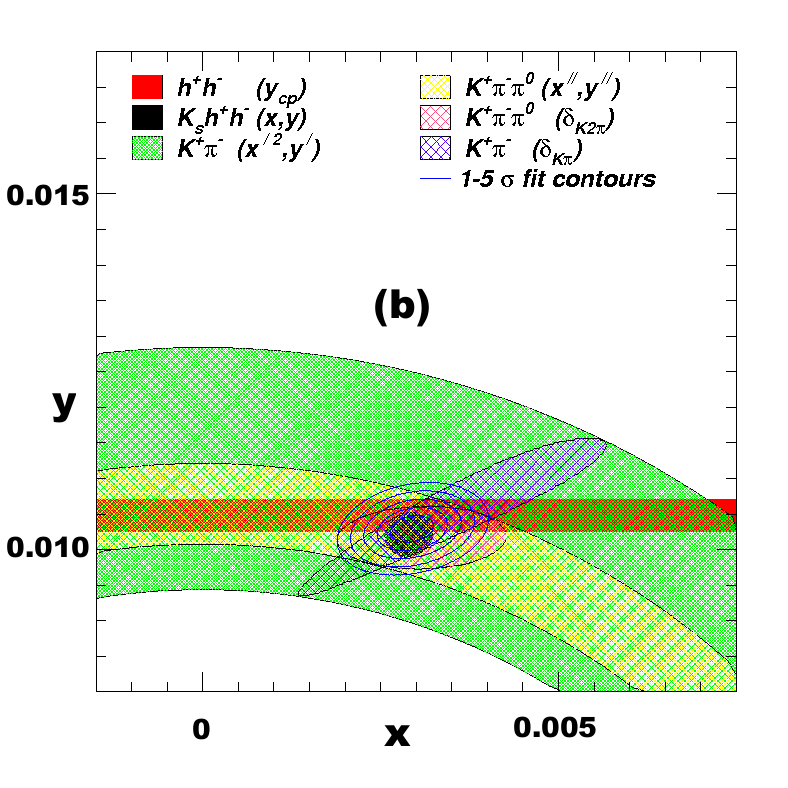}
}
\end{tabular}
  \caption{Mixing observables from \babar\ projected into the (\xd,\yd)
   plane.  Shaded areas indicate the 68.3\% confidence region for each.
   Contours enclosing $1-5\sigma$ two-dimensional confidence regions
   from $\chi^2$ fits to these observables are drawn as solid blue lines.
   (a) Includes results anticipated from scaling \babar\ results by
   a factor 100 in statistical significance anticipated for \superb.
   Uncertainties in $\xd (\pm 7.5\times 10^{-4}$) and
   $\yd (\pm 1.9\times 10^{-4})$, are dominated by uncertainty in the
   amplitude model.
   In (b), projections of hypothetical measurements of $\delta_{\Kp\pim}$
   and $\delta_{\Kp\pim\piz}$ are included and are expected to reduce
   the amplitude model uncertainty by a factor 10 leading to uncertainties
   in $\xd (\pm 2.0\times 10^{-4}$) and $\yd (\pm 1.2\times 10^{-4})$.
  }
  \label{fig:bmfig}
  \end{center}
\end{figure}


\section{\label{sec:summary}Summary}

In conclusion, there is strong evidence for mixing from a variety of measurements, but no one observation at the $5\sigma$ level has yet been made.  No evidence has yet been seen for \CPV.  Prospects are that \superb\ or Belle II will improve precision in \xd\ and \yd\ to the $10^{-4}$ level providing sensitivity to \CPV\ ($|q/p|$) at about $2-5$\% from channel to channel.  However, this will only be possible if either new measurements of strong phases, or new amplitude analysis techniques become available.


\end{document}